\documentclass[a4paper, conference]{IEEEtran}
\IEEEoverridecommandlockouts
\usepackage[utf8]{inputenc}
\usepackage{amsmath,amsfonts}
\usepackage{algorithm}
\usepackage{algpseudocode}
\usepackage{booktabs} 
\usepackage{array}    
\usepackage{tabularx}
\usepackage[caption=false,font=normalsize,labelfont=sf,textfont=sf]{subfig}
\usepackage{textcomp}
\usepackage{stfloats}
\usepackage{url}
\usepackage{verbatim}
\usepackage{graphicx}
\usepackage{cite}
\usepackage{graphicx}
\usepackage{tikz}
\usepackage{pgfplots}
\usepackage{psfrag}
\usepackage[justification=centering]{caption}
\usetikzlibrary{arrows,snakes,shapes,shapes.geometric,fit,positioning,arrows,decorations.markings,calc}
\usepackage{mathtools}
\usepackage{amsbsy}
\usepackage{fixmath}
\usepackage{amssymb}
\usepackage{mathrsfs}
\usepackage{fancyhdr}
\usepackage{xcolor}
\usepackage{hyperref}

\newcommand{\HH}{\mathrm{H}}

\newcommand{\tdl}{\mathrm{dl}}
\newcommand{\bs}[1]{\boldsymbol{#1}}

\newcommand{\tr}{\text{tr}}
\newcommand*{\rttensor}[1]{\overline{\overline{#1}}}

\definecolor{lime}{HTML}{A6CE39}
\DeclareRobustCommand{\orcidicon}{%
	\begin{tikzpicture}
	\draw[lime, fill=lime] (0,0) 
	circle [radius=0.16] 
	node[white] {{\fontfamily{qag}\selectfont \tiny ID}};
	\draw[white, fill=white] (-0.0625,0.095) 
	circle [radius=0.007];
	\end{tikzpicture}
	\hspace{-2mm}
}

\foreach \x in {A, ..., Z}{%
	\expandafter\xdef\csname orcid\x\endcsname{\noexpand\href{https://orcid.org/\csname orcidauthor\x\endcsname}{\noexpand\orcidicon}}
}

\makeatletter
\def\endthebibliography{%
	\def\@noitemerr{\@latex@warning{Empty `thebibliography' environment}}%
	\endlist
}
\makeatother

\begin{document}
	
	\title{Highly Accelerated Weighted MMSE Algorithms for Designing Precoders in FDD Systems with Incomplete CSI 
	}
	
	\author{\IEEEauthorblockN{Donia~Ben~Amor\orcidA{}, Michael~Joham\orcidB{}, Wolfgang~Utschick\orcidC{}}
		\IEEEauthorblockA{ 	\textit{Chair of Methods of Signal Processing} \\
			\textit{School of Computation, Information and Technology, Technical University of Munich} \\
			Munich, Germany \\
			Email: \{donia.ben-amor, joham, utschick\}@tum.de}
	}
	\maketitle

	\begin{abstract}
In this work, we derive a lower bound on the training-based achievable downlink (DL) sum rate (SR) of a multi-user multiple-input-single-output (MISO) system operating in frequency-division-duplex (FDD) mode. Assuming linear minimum mean square error (LMMSE) channel estimation is used, we establish a connection of the derived lower bound on the signal-to-interference-noise-ratio (SINR) to an average MSE that allows to reformulate the SR maximization problem as the minimization
of the augmented weighted average MSE (AWAMSE). We propose an iterative precoder design with three alternating steps, all given in closed form, drastically reducing the computation time. We show numerically the effectiveness of the proposed approach in challenging scenarios with limited channel
knowledge, i.e., we consider scenarios with a very limited number of pilots. We additionally propose a more efficient version of the well-known stochastic iterative WMMSE (SIWMMSE) approach, where the precoder update is given in closed form.
	\end{abstract}
	
	
	\section{Introduction}
	Several works on precoder design for multiple-input-multiple-output (MIMO) systems mainly consider time-division-duplex (TDD) mode since it offers the advantage of reciprocity between the uplink (UL) and the downlink (DL) channels \cite{Tse}. Frequency-division-duplex (FDD) systems are, however, more challenging since this reciprocity does not hold due to the frequency gap between UL and DL carrier frequencies \cite{Tse}. Therefore, the base station (BS) has to send known pilots to the users, which in turn feed back the channel observations to the BS to allow estimating the DL channels at the BS and designing transmit strategies accordingly. Note that if complete channel state information (CSI) is desired, the number of pilots should be at least equal to the number of BS antennas.  Due to the tendency of deploying more and more antennas at the BS, also known as massive MIMO \cite{marzetta}, and considering the limited channel coherence interval, i.e., the time for which the channel can be assumed to be constant, no room is left for data transmission by having long training phases. Therefore, we focus on the case where only a few channel observations are available. In order to cope with this incomplete channel knowledge, when designing transmit strategies at the BS, we derive a lower bound on the training-based achievable sum rate (SR) that takes into account the channel estimation errors \cite{Hasssibi_Hochwald} and, hence, does not use naively the available CSI as if it were perfect. We shall note that this lower bound was used for the UL in \cite{MMSEprecoder} to derive the MMSE equalizer, and by the means of some UL-DL duality, the MMSE precoder is found. In this work, we derive the lower bound for the DL, i.e., from the BS perspective, and therefore, it is directly suitable for precoder design. Furthermore, we define an average mean square error (AMSE) of the data signals, where the averaging is with respect to the channel estimates. By introducing receive filters at the users and weights for each AMSE, we establish a relationship of the augmented weighted AMSE (AWAMSE) to the derived lower bound on the rate. This allows us to recast the SR maximization as the minimization of the sum of AWAMSEs. The latter optimization problem, though it is non-convex, is convex in each of its three variables, namely the receive filters, the weights, and the precoding matrix when the other two are fixed. The two former variables are directly given in closed form, whereas the precoder update is the most costly step, and interior-point methods or line-search approaches are usually applied to find the optimal precoding matrix \cite{IWMMSE_Luo}. To avoid associated high complexity, we propose to reformulate the precoder optimization problem as an unconstrained one, such that a closed-form solution for the precoding matrix in each iteration step is obtained and only a re-scaling of the precoder is needed after the algorithm has converged. This equivalent formulation of the optimization problem is based on introducing a common scaling at the receivers and at the transmitter, as proposed in \cite{WienerFilter}. Inspired by this idea, works like \cite{WSRmax}, \cite{RethinkWMMSE}, \cite{UnfoldingIWMMSE} proposed to modify the standard iterative weighted minimum mean square error (IWMMSE) algorithm \cite{IWMMSE_Luo} such that the precoder update is given in closed form and line-search methods like bisection are avoided. By adapting this idea to our setup, we end up with an iterative approach where each variable update is given in closed form, and, hence, the computation time is drastically reduced. \\
	Additionally, we review the stochastic IWMMSE (SIWMMSE) algorithm \cite{SWMMSE} and propose to adapt the precoder update step, such that it is given in closed form. Consequently, The computational complexity of the SIWMMSE remains in the Monte Carlo runs inside each iteration that are used to average over all the channel samples.
	
	\par This paper is organized as follows: In Section~\ref{sec:SysModel}, we present the system model for the considered setup and derive our figure-of-merit in Section~\ref{sec:Fig-of-Merit}. In Section~\ref{sec:PrecDesign}, we present our precoding approach and later in Section \ref{sec:sIWMMSE}, we review the SIWMMSE algorithm and propose a more efficient version. The performance of the different precoding approaches is then evaluated through numerical results in Section~\ref{sec:results}.

\section{System Model}\label{sec:SysModel}
We consider the DL of a multi-user MISO FDD system, where the BS is equipped with $M$ antennas and serves $K$ single-antenna users. The channel between user $k$ and the BS is denoted by $\bs{h}_k \sim \mathcal{N}_\mathbb{C}(\bs{0},\bs{C}_k)$.
During the DL probing phase, user $k$ receives $T_\tdl$ channel observations which are collected in $\bs{y}_k=\bs{\Phi}^\HH \bs{h}_k + \bs{z}_k$, 
where $\bs{z}_k \sim \mathcal{N}_\mathbb{C}(\bs{0}, \frac{1}{P_\tdl} \bs{I})$ denotes the normalized DL training noise, $\bs{\Phi}\in \mathbb{C}^{M\times T_\tdl}$ is the pilot matrix and $P_\tdl$ denotes the DL transmit power available at the BS.
For the sake of simplicity, we assume that all users send their channel observations interference-free via analog feedback to the BS. \\
During the data transmission phase, the information carrying symbols $s_k\sim \mathcal{N}_\mathbb{C}(0,1), k=1,\dots K$, are precoded and transmitted to the users. The signal received at user $k$ is given by
\begin{equation}
   r_k  = \bs{h}_k^\HH \bs{p}_k s_k + \sum_{j\neq k} \bs{h}_k^\HH \bs{p}_j s_j + n_k.
\end{equation}
The $n_k \sim \mathcal{N}_\mathbb{C}(0,1)$ denotes the AWGN at the $k$th user's antenna and $\bs{p}_k$ is the precoder for the $k$th user. All users' precoders are subject to the transmit power constraint $\sum_k \|\bs{p}_k\|^2 \leq P_\tdl$.\\
The achievable instantaneous rate of user $k$ for perfect receiver CSI can be written as
\begin{equation}
    R_k=\log_2\left(1+\frac{|\bs{h}_k^\HH \bs{p}_k|^2}{\sum_{j\neq k}|\bs{h}_k^\HH \bs{p}_j|^2+1}\right). \label{eq:instRk}
\end{equation}
Due to the incomplete CSI at the transmitter side, relying on \eqref{eq:instRk} for precoder design can lead to poor performance since using the channel estimate as if it were the true channel can be detrimental, as shown in Section \ref{sec:results}. Therefore, we model the $k$th user's channel as \cite{Hasssibi_Hochwald}
\begin{equation}
    \bs{h}_k= \hat{\bs{h}}_k + \Tilde{\bs{h}}_k \label{eq:hk}
\end{equation}
where the channel estimate $\hat{\bs{h}}_k$ is considered to be known at the BS.  $\Tilde{\bs{h}}_k$ is the zero-mean estimation error whose covariance matrix is denoted by $\bs{C}_{\text{err},k}$. The received signal can be thus rewritten as
\begin{equation}
      r_k=  \underbrace{\hat{\bs{h}}_k^\HH \bs{p}_k}_{h_{\text{eff},k}} s_k + \underbrace{\Tilde{\bs{h}}_k^\HH \bs{p}_k s_k +  \sum_{j\neq k} \bs{h}_k^\HH \bs{p}_j s_j}_{v_{k}} + n_k.\label{eq:rk_l2}  
\end{equation}
Next, we derive a lower bound on the achievable rate for the underlying DL setup using the model defined in \eqref{eq:hk}. This bound is then used for the precoder optimization at the transmitter.

\section{Figure-of-Merit}\label{sec:Fig-of-Merit}
We aim to formulate a lower bound on the training-based SINR of user $k$ using the model in \eqref{eq:rk_l2} corresponding to a discrete memoryless interference channel with input $s_k$, output $r_k$, independent noise $n_k$, random interference $v_k$ and known channel response $h_{\text{eff},k}$. We additionally denote by $u=\{\hat{\bs{h}}_i\}$ the realizations of all users' channel estimates. Note that $h_{\text{eff},k}$ and $u$ are random but assumed to be fixed during the channel coherence block. \\
The ergodic capacity $C_k$ of such a channel is lower bounded as follows \cite{massiveMIMObook}
\begin{equation}
    C_k \geq \rttensor{R}_k = \mathbb{E}\left[ \log_2\left( 1 + \rttensor{\text{SINR}}_k \right) \right] \label{eq:ergC}
\end{equation}
with 
\begin{equation}
    \rttensor{\text{SINR}}_k = \frac{|h_{\text{eff},k}|^2}{\mathbb{E}\left[|v_k|^2| u\right]+1}\label{eq:SINR1}
\end{equation}
where the interference term $v_k$ has conditionally zero mean, i.e., $\mathbb{E}[v_k|u]=0$ and is conditionally uncorrelated with the input signal $s_k$, i.e., $\mathbb{E}[v_k s_k^*|u]=0$. The latter holds under the assumption that the estimation error is independent of the channel estimate.\\
Assuming that LMMSE channel estimation is performed, the estimation error $\Tilde{\bs{h}}_k$ and the channel estimate $\hat{\bs{h}}_k$ are uncorrelated due to the orthogonality principle \cite{Scharf1989}.
The LMMSE channel estimate of the $k$th user is given by \cite[Section 8.3]{Scharf1989}
\begin{equation}
    \hat{\bs{h}}_k^\text{LMMSE}=\bs{C}_k\bs{\Phi}\left(\bs{\Phi}^\HH \bs{C}_k \bs{\Phi} + \sigma^2 \bs{I}\right)^{-1}\bs{y}_k.
\end{equation}
Now, recalling that $\bs{C}_{\text{err},k}=\mathbb{E}\left[\Tilde{\bs{h}}_k\Tilde{\bs{h}}_k^\HH\right]$ is the covariance matrix of the estimation error, the expectation in \eqref{eq:SINR1} can be evaluated as follows
\begin{align}
    &\mathbb{E}\left[|v_k|^2| u\right]= \mathbb{E}\left[|v_k|^2| \{\hat{\bs{h}}_i\}\right] \\ 
    &= \bs{p}_k^\HH \bs{C}_{\text{err},k}  \bs{p}_k +  \sum_{j\neq k} \left( \bs{p}_j^\HH \hat{\bs{h}}_k\hat{\bs{h}}_k^\HH \bs{p}_j + \bs{p}_j^\HH  \bs{C}_{\text{err},k} \bs{p}_j\right) \\
    &=\sum_{j=1}^K  \bs{p}_j^\HH  \bs{C}_{\text{err},k} \bs{p}_j +\sum_{j\neq k} |\hat{\bs{h}}_k^\HH \bs{p}_j|^2
\end{align}
where we used the unit variance of the input signals and the independence of the estimation error from the channel estimate.\\
The effective SINR is, therefore, given by
\begin{equation}
     \rttensor{\text{SINR}}_k = \frac{|\hat{\bs{h}}_k^\HH \bs{p}_k|^2}{\sum_{j=1}^K  \bs{p}_j^\HH  \bs{C}_{\text{err},k} \bs{p}_j +\sum_{j\neq k} |\hat{\bs{h}}_k^\HH \bs{p}_j|^2+1}.\label{eq:SINR2}
\end{equation}

In order to design the precoding vectors $\bs{p}_1, \dots, \bs{p}_K$, we use an average MSE of the data signals as a figure-of-merit, which is closely related to the effective SINR given by \eqref{eq:SINR2}. \\
To this end, we first define $g_k$ to be the receive filter at user $k$. The estimate of the data signal $s_k$ thus reads as
\begin{equation}
    \hat{s}_k=g_k r_k= g_k \hat{\bs{h}}_k^\HH \bs{p}_k s_k + g_k \Tilde{\bs{h}}_k^\HH \bs{p}_k s_k + g_k \sum_{j\neq k} \bs{h}_k^\HH \bs{p}_j s_j + g_k n_k. \label{eq:shat}
\end{equation}
The MSE of the data signal $s_k$ is given by
$\varepsilon_k = \mathbb{E}\left[ |\hat{s}_k - s_k |^2\right]$,
where the expectation is w.r.t. the data signals and the noise signal.
Using \eqref{eq:shat}, the MSE can be evaluated as follows
\begin{align*}
    \varepsilon_k &= \mathbb{E}\left[ |\hat{s}_k - s_k |^2\right]\\
   &=|g_k|^2 |\hat{\bs{h}}_k^\HH \bs{p}_k |^2 +|g_k|^2 |\Tilde{\bs{h}}_k^\HH \bs{p}_k |^2+  |g_k|^2  \bs{p}_k^\HH \Tilde{\bs{h}}_k \hat{\bs{h}}_k^\HH \bs{p}_k \\&+ |g_k|^2  \bs{p}_k^\HH \hat{\bs{h}}_k \Tilde{\bs{h}}_k^\HH \bs{p}_k  -2\Re\{g_k \hat{\bs{h}}_k^\HH \bs{p}_k \} -2\Re\{g_k \Tilde{\bs{h}}_k^\HH \bs{p}_k \} \\&+ |g_k|^2 \sum_{j\neq k} |\bs{h}_k^\HH \bs{p}_j |^2  + |g_k|^2 +1.
\end{align*}
We now define an average MSE w.r.t. the channel estimates $\rttensor{\varepsilon}_k=\mathbb{E}[\varepsilon_k]$. Due to the independence of the channel estimate $\hat{\bs{h}}_k$ and the estimation error $\Tilde{\bs{h}}_k$, $\rttensor{\varepsilon}_k$ is given by
\begin{equation}\label{eq:AvEps}
 \begin{aligned} 
   \rttensor{\varepsilon}_k&=|g_k|^2 \sum_j  |\hat{\bs{h}}_k^\HH \bs{p}_j|^2 + |g_k|^2 \sum_j \bs{p}_j^\HH \bs{C}_{\text{err},k} \bs{p}_j \\
    &- 2\Re\{g_k \hat{\bs{h}}_k^\HH \bs{p}_k\} +|g_k|^2 +1
\end{aligned}   
\end{equation}
where we used the fact that the estimation error is zero-mean and has the covariance matrix $\bs{C}_{\text{err},k}$.\\
The MSE minimizing receive filter hence reads as
\begin{equation} \label{eq:gkMMSE}
    g_k^\text{MMSE} = \bs{p}_k^\HH \hat{\bs{h}}_k T_k^{-1}.
\end{equation}
with $T_k=\sum_j |\hat{\bs{h}}_k^\HH \bs{p}_j|^2 + \bs{p}_j^\HH \bs{C}_{\text{err},k} \bs{p}_j + 1$. 
Inserting this expression into \eqref{eq:AvEps} results in
$ \rttensor{\varepsilon}_k^\text{MMSE} = 1 -  |\hat{\bs{h}}_k^\HH \bs{p}_k|^2 T_k^{-1}$.\\
Note that $\rttensor{\varepsilon}_k^\text{MMSE}$ is related to the rate lower bound $\rttensor{R}_k$ of user $k$ defined in \eqref{eq:ergC} according to
$\rttensor{R}_k= \log_2(1+\rttensor{\text{SINR}}_k)=-\log_2( \rttensor{\varepsilon}_k^\text{MMSE})$.
Now, we introduce the augmented weighted average MSE (AWAMSE) of user $k$ defined as
\begin{equation}
    \rttensor{\xi}_k=u_k \rttensor{\varepsilon}_k - \log_2 u_k
\end{equation}
where $u_k$ is the weight associated with the $k$th user's MSE. It can be shown that minimizing $ \rttensor{\xi}_k$ w.r.t. the receive filters $g_k$ and the weight $u_k$ yields the following relationship to the rate lower bound
$\rttensor{\xi}_k^\text{MMSE}=1-\rttensor{R}_k$,
where the optimal receive filter is the MMSE receive filter given by \eqref{eq:gkMMSE} and the optimal weight reads as $u_k^\text{MMSE}=1/\rttensor{\varepsilon}_k^\text{MMSE}$.

Establishing the connection of the AWAMSE to the rate bound allows us to unveil hidden convexity properties of the SR maximization problem. For instance, by fixing the weights and the receive filters, optimizing the precoders turns out to be a convex optimization problem that is usually solved using off-the-shelf solvers. In the following, we show how the usage of convex solvers can even be circumvented and how a closed-form expression for the precoder can be found.	

\section{Precoder Design}\label{sec:PrecDesign}
First, we start by formulating the underlying optimization stemming from $\underset{\bs{P},\bs{U},\bs{G}}{\min}\,\sum_{k} \rttensor{\xi}_k$ in the following compact form 
\begin{equation}
  \begin{aligned}
    \underset{\bs{P}, \bs{U},\bs{G}}{\min}\: &\tr(\bs{U})  - 2 \Re\{\tr (\bs{U} \bs{G} \hat{\bs{H}}^\HH\bs{P}) \}+
    \tr (\bs{U} \bs{G} \hat{\bs{H}}^\HH \bs{P}\bs{P}^\HH
    \hat{\bs{H}}\bs{G}^\ast) \\    + & \tr (\bs{P}\bs{P}^\HH \bs{Z}) + \tr (\bs{U}\bs{G}\bs{G}^\ast) - \log_2(\det(\bs{U}))\\
    \text{s.t.} \: &\|\bs{P}\|_\mathrm{F}^2\leq P_\tdl
\end{aligned}  \label{eq:AWAMSEmin2}
\end{equation}
where $\bs{P}=[\bs{p}_1, \dots, \bs{p}_K]$, $\bs{U}=\text{diag}(u_1, \dots, u_K)$, and $\bs{G}=\text{diag}(g_1, \dots, g_K)$. We also introduced $\bs{Z}=\sum_k u_k |g_k|^2 \bs{C}_{\text{err},k}$ and the estimated channel $\hat{\bs{H}}=[\hat{\bs{h}}_1,\dots, \hat{\bs{h}}_K]$.

Before proceeding to solve the optimization problem in \eqref{eq:AWAMSEmin2}, we introduce a common scaling $\beta^{-1}$ of the received signals $\hat{s}_k$, which allows their amplitude to be different than the input signals $s_k$ \cite{WienerFilter}. Thus, our new MSE reads as $\mathrm{E}[|s_k-\beta^{-1}\hat{s}_k|^2]$. Thus, the optimization problem in \eqref{eq:AWAMSEmin2} can be rewritten as
\begin{equation}
  \begin{aligned}
   & \underset{\bs{P}, \bs{U},\bs{G}, \beta}{\min}\: \tr(\bs{U})  - 2 \beta^{-1}\Re\{\tr (\bs{U} \bs{T}\bs{P}) \}+  \beta^{-2}\tr (\bs{P}\bs{P}^\HH \bs{Z})+\\&
    \beta^{-2}\tr (\bs{U} \bs{T}\bs{P}\bs{P}^\HH
   \bs{T}^\HH)      + \beta^{-2}\tr (\bs{U}\bs{G}\bs{G}^\ast) - \log_2(\det(\bs{U}))\\
    &\text{s.t.} \quad \tr (\bs{P}\bs{P}^\HH) \leq P_\tdl
\end{aligned}  \label{eq:AWAMSEbeta}
\end{equation}
where we replaced $\bs{G} \hat{\bs{H}}^\HH$ by $\bs{T}$.
Denoting by $\lambda$ the Lagrangian multiplier associated with the power constraint in \eqref{eq:AWAMSEbeta}, we can formulate the Lagrangian function corresponding to \eqref{eq:AWAMSEbeta} as follows
\begin{equation}
\begin{aligned}
 \mathcal{L}&=\tr(\bs{U})  - 2 \beta^{-1}\Re\{\tr (\bs{U} \bs{T} \bs{P}) \}+  \beta^{-2}\tr (\bs{P}\bs{P}^\HH \bs{Z}) \\ 
&+\beta^{-2}\tr (\bs{U} \bs{T} \bs{P}\bs{P}^\HH
\bs{T}^\HH)      + \beta^{-2}\tr (\bs{U}\bs{G}\bs{G}^\ast) \\
&- \log_2(\det(\bs{U}) +\lambda (\tr(\bs{P}\bs{P}^\HH)-P_\tdl )
\end{aligned}
\end{equation}
For fixed $\bs{U}$, $\bs{G}$ and $\beta$, the precoding matrix is given in closed form using the first-order optimality condition
\begin{equation}
    \bs{P}=\beta \left(\bs{Z}+\bs{T}^\HH \bs{U} \bs{T} + \delta \bs{I}\right)^{-1} \bs{T}^\HH\bs{U}
\end{equation}
where $\delta=\beta^2 \lambda$. From the power constraint, we can obtain an expression for $\beta$ depending on $\delta$
\begin{equation}\label{eq:beta}
    \beta(\delta)=\sqrt{\frac{P_\tdl}{\tr\left((\bs{Z}+\bs{T}^\HH \bs{U} \bs{T} + \delta \bs{I})^{-2}\bs{T}^\HH\bs{U}\bs{U}\bs{T}\right)}}.
\end{equation}
By this choice of $\beta$, we guarantee that the transmit power constraint is satisfied with equality. The constrained optimization problem in \eqref{eq:AWAMSEbeta} can accordingly be recast as the following unconstrained optimization problem w.r.t. $\delta$ 
\begin{equation}
\begin{aligned}
       \underset{\delta}{\min}\: &\tr(\bs{U})  - 2 \Re\{\tr (\bs{U}\bs{T}\bs{A}(\delta)^{-1}\bs{T}^\HH \bs{U}) \}\\
       +&\tr (\bs{A}(\delta)^{-1}\bs{T}^\HH \bs{U}\bs{U} \bs{T} \bs{A}(\delta)^{-1} \bs{Z}) \\
   + & \tr (\bs{U} \bs{T} \bs{A}(\delta)^{-1} \bs{T}^\HH \bs{U}\bs{U} \bs{T} \bs{A}(\delta)^{-1} \bs{T}^\HH) \\      + &\frac{1}{P_\tdl}\tr(\bs{A}(\delta)^{-2}\bs{T}^\HH \bs{U}\bs{U} \bs{T}) \tr (\bs{U}\bs{G}\bs{G}^\ast) - \log_2(\det(\bs{U}))
\end{aligned} \label{eq:AWAMSEmin3}
\end{equation}
where we defined $\bs{A}(\delta)=\bs{Z}+\bs{T}^\HH \bs{U} \bs{T} + \delta \bs{I}$. The optimal $\delta^\text{opt}$ is found by setting the first derivative of the objective in \eqref{eq:AWAMSEmin3} to zero and reads as
\begin{equation}\label{eq:DeltaOpt}
    \delta^\text{opt}=\tr(\bs{U}\bs{G}\bs{G}^\star)/P_\tdl.
\end{equation}
The derivation of $\delta^\text{opt}$ is given in the Appendix (see Section~\ref{Appendix}).
\\
Hence, the unconstrained precoding matrix $\bs{P}^\text{unconst}$ is given by
\begin{equation}
    \bs{P}^\text{unconst}= \left(\bs{Z} + \hat{\bs{H}}\bs{G}^\ast \bs{U} \bs{G} \hat{\bs{H}}^\HH + \frac{\tr(\bs{U}\bs{G}\bs{G}^\ast)}{P_\tdl} \bs{I} \right)^{-1} \hat{\bs{H}} \bs{G}^\ast \bs{U}.\label{eq:Pk}
\end{equation}
The receive filters $g_k$ and the weights $u_k$ are computed based on the objective in \eqref{eq:AWAMSEmin3} by fixing the other variables, that is
\begin{align}
    g_k&= \bs{p}_k^\HH \hat{\bs{h}}_k \Tilde{T}_k^{-1} \label{eq:gk_scaled}\\
    u_k&=(1 - |\hat{\bs{h}}_k^\HH \bs{p}_k|^2\Tilde{T}_k^{-1})^{-1}\label{eq:uk_scaled}
\end{align}
with $\bs{p}_k$ is the $k$th column of $\bs{P}^\text{unconst}$ and $\Tilde{T}_k=\sum_j |\hat{\bs{h}}_k^\HH \bs{p}_j|^2 + \bs{p}_j^\HH \bs{C}_{\text{err},k} \bs{p}_j + \|\bs{p}_j\|^2/P_\tdl$.\\
The AWAMSE algorithm used to solve \eqref{eq:AWAMSEmin3} is summarized in Algorithm~\ref{alg:myalgorithm}.
\begin{algorithm}
\caption{AWAMSE Algorithm}
\label{alg:myalgorithm}
\begin{algorithmic}[1] 
\State  Initialize the precoding matrix $\bs{P}$
\State Compute the receive filters $g_k$ according to \eqref{eq:gk_scaled}
\State Compute the weighting factors $u_k$ according to \eqref{eq:uk_scaled}
\State Calculate the precoding matrix $\bs{P}^\text{unconst}$ based on \eqref{eq:Pk}
\State Repeat steps 2-4 until convergence of the sum rate lower bound
\State Scale the precoding matrix with $\beta$ [see \eqref{eq:beta}] to satisfy the power constraint
\end{algorithmic}
\end{algorithm}

\section{Stochastic IWMMSE} \label{sec:sIWMMSE}
In this section, we briefly review the SIWMMSE approach \cite{SWMMSE}, \cite{Clerckx_RS3} and propose a closed-form update of the precoders that circumvents using interior-point methods and, therefore, accelerates the convergence of the algorithm. The SIWMMSE method is based on averaging over channel samples that can be generated using channel statistics. In contrast to \cite{Clerckx_RS3}, we assume correlated and not i.i.d. channels and, thus, using the sampling technique proposed there leads to poor performance. Hence, we suggest to draw the $n$th channel sample of user $k$ according to 
\begin{equation}\label{eq:sampleGen}
\bs{h}_k^{(n)}=\hat{\bs{h}}_k + \bs{C}_{\text{err},k}^{\frac{1}{2}} \Tilde{\bs{e}}_k
\end{equation}
where $\hat{\bs{h}}_k$ is the channel estimate, $\Tilde{\bs{e}}_k $ is generated at random from the normal Gaussian distribution, i.e., $\Tilde{\bs{e}}_k \sim \mathcal{N}_\mathbb{C}(\bs{0},\bs{I})$, and $\bs{C}_{\text{err},k}^{\frac{1}{2}}$ denotes the square root matrix of $\bs{C}_{\text{err},k}$.

The SIWMMSE algorithm aims at solving the stochastic average rate maximization problem given by
\begin{equation}\label{eq:ARmax}
    \underset{\bs{P}}{\max}\, \sum_{k=1}^K \overline{R}_k \quad \text{s.t.} \quad \|\bs{P}\|_\mathrm{F}^2 \leq P_\tdl
\end{equation}
where the average rate of user $k$ is a performance measure over the error distribution for a given channel estimate, i.e., $\overline{R}_k=\mathbb{E}_{\bs{H}|\hat{\bs{H}}}[R_k|\hat{\bs{H}}]$. In order to make \eqref{eq:ARmax} tractable, the sample average approximation is used to approximate the average rates in the first step, then a relationship between the approximated rate and the weighted MSE is established to unveil hidden convexity properties of the underlying problem as presented in Section~\ref{sec:Fig-of-Merit}.\\
The approximated average rate is given by $\overline{R}_k^{(N)}=\frac{1}{N}\sum_{n=1}^N R_k^{(n)}$, where $R_k^{(n)}=R_k(\bs{H}^{(n)})$ is the instantaneous rate [cf. \eqref{eq:instRk}] evaluated for the $n$th channel sample. By introducing receive filters $g_k$ and weights $u_k$ for each user $k$, one can define an approximated weighted MSE per user that is closely related to $\overline{R}_k^{(N)}$ \cite{Clerckx_RS3}. \\
In the $i$th iteration of the SIWMMSE algorithm, for a fixed precoding matrix, a set of variables is computed for the $n$th channel realization
\begin{align*}
    t_k^{(n)}=u_k^{(n)}|g_k^{(n)}|^2, \quad &v_k^{(n)}=\log_2(u_k^{(n)})\\
    \bs{f}_k^{(n)}=u_k^{(n)}g_k^{(n),*} \bs{h}_k^{(n)}, \quad &\bs{\Psi}_k^{(n)}=t_k^{(n)}\bs{h}_k^{(n)} \bs{h}_k^{(n),\HH}
    \end{align*}
In the second step, and considering the average of the previously computed variables over the $N$ channel samples, the precoding matrix is updated by solving the following optimization problem
\begin{equation}\label{eq:SWMMSEconst}
\begin{aligned}
     &\underset{\bs{P}}{\min} \: \sum_{k=1}^K\sum_{j=1}^K \bs{p}_j^\HH \overline{\bs{\Psi}}_k \bs{p}_j +\overline{t}_k - 2\Re\{\overline{\bs{f}}_k^\HH \bs{p}_k\} +\overline{u}_k-\overline{v}_k\\
     &\text{s.t.} \quad \|\bs{P}\|_\mathrm{F}^2 \leq P_\tdl
\end{aligned}
\end{equation}
where $\overline{a}=\frac{1}{N}\sum_{n=1}^N a^{(n)}$ denotes the average of some variable $a$ over the $N$ channel realizations. 
\eqref{eq:SWMMSEconst} is convex in the precoding matrix $\bs{P}$ and interior-point methods are usually used to solve it. Alternatively, the precoder can be found by formulating the Lagrangian function corresponding to \eqref{eq:SWMMSEconst} and defining $\lambda$ to be the Lagrangian multiplier associated with the power constraint. A line search (e.g., via bisection) for the optimal Lagrangian multiplier has then to be performed, and the precoder is given, therefore, in quasi-closed form.\\
In order to avoid any line search, and similarly to the discussion in the previous section, we can recast \eqref{eq:SWMMSEconst} as an unconstrained optimization problem as follows 
\begin{equation}
    \underset{\bs{P}}{\min} \: \sum_{k=1}^K\sum_{j=1}^K \bs{p}_j^\HH \overline{\bs{\Psi}}_k \bs{p}_j +\frac{\tr(\bs{P}\bs{P}^\HH)}{P_\tdl}\overline{t}_k - 2\Re\{\overline{\bs{f}}_k^\HH \bs{p}_k\} +\overline{u}_k-\overline{v}_k.
\end{equation}
The precoder update is therefore given in closed form as
\begin{equation}
    \bs{P}^\text{unconst}=\left(\sum_{j=1}^K\overline{\bs{\Psi}}_j + \frac{1}{P_\tdl}\sum_{j=1}^K  \overline{t}_j \bs{I} \right)^{-1}\overline{\bs{F}}_k
\end{equation}
where $\overline{\bs{F}}_k=[\overline{\bs{f}}_1, \dots, \overline{\bs{f}}_K]$. 
After the algorithm has converged, the precoding matrix has to be scaled to satisfy the power constraint (as in line 6 of Algorithm~\ref{alg:myalgorithm}).
 
\section{Results}\label{sec:results}
Throughout our simulations, we consider $T_\tdl<M$ so that the channel knowledge is incomplete at the BS. We assume that the BS performs linear MMSE channel estimation in order to match our assumptions for the lower bound derived in Section~\ref{sec:Fig-of-Merit}.
We generate the channels according to $\bs{h}_k\sim\mathcal{N}_\mathbb{C}(\bs{0},\bs{C}_k)$, where $\bs{C}_k$ is the covariance matrix corresponding to a Gaussian mixture model (GMM) \cite{GMM2} component obtained by fitting a GMM to the DL training observations of channels generated using QUAsi Deterministic RadIo channel GenerAtor (QuaDRiGa) \cite{quadriga}.
Unless otherwise specified, all the results are averaged over $100$ different setups corresponding to different random selections of the covariance matrices. For each setup, we consider $100$~channel realizations. For all the iterative approaches, we use the same termination criterion and set the maximum number of iterations to $100$.

 In the following, we initialize all the algorithms with the MMSE precoder \cite{MMSEprecoder} given by $\bs{P}^\text{mmse}=\beta \left(\hat{\bs{H}}\hat{\bs{H}}^\HH+\bs{C}_\text{err}+\alpha\bs{I}\right)^{-1}\hat{\bs{H}}$, where $\bs{C}_\text{err}=\sum_k \bs{C}_{\text{err},k}$. The normalization factor $\beta$ is used to satisfy the power constraint, and $\alpha=M/P_\tdl$ is a regularization parameter. 

We first present the results obtained by directly optimizing the instantaneous SR [cf. \eqref{eq:instRk}] using the standard IWMMSE approach, where the filters, the weights, and the precoding matrix are updated in an alternating fashion. Here, we also circumvent any line search needed for the precoder update by introducing a re-scaling approach similar to the one presented in Section~\ref{sec:PrecDesign}. Note that the standard IWMMSE algorithm treats the channel estimates as if they were the true channels and is, hence, unaware of any CSI errors. This can lead to an overestimation of the achievable rates and, therefore to poor performance, as illustrated in Fig.\ref{fig:Obj_vs_True_IWMMSE}. Here, we evaluate the SR achieved for $M=32$ antennas, $K=8$ users, and $T_\tdl=4$ pilots, before and after optimization based on both the estimated and the true channel at a transmit power of $40$~dB. One can see in Fig.\ref{fig:Obj_vs_True_IWMMSE} that the optimization using the standard IWMMSE yields a significant improvement of the SR evaluated based on the estimated channel, whereas the actual achievable SR exhibits a decrease after optimizing the precoders due to the mismatch between the true and estimated channel.
\begin{figure}[h!]
\begin{minipage}{0.48\columnwidth}
\centering
\scalebox{0.4}{
%
%
\definecolor{mycolor1}{rgb}{0.00000,0.44700,0.74100}%
\definecolor{mycolor2}{rgb}{0.85000,0.32500,0.09800}%
\begin{tikzpicture}

\begin{axis}[%
width=3.521in,
height=3in,
at={(0.758in,0.481in)},
scale only axis,
ybar=2.5pt,
bar width=.8cm,
xmin=0.514285714285714,
xmax=2.48571428571429,
xtick={1,2},
yticklabel style={font=\Large},
xticklabel style={font=\Huge},
xticklabels={{$\substack{\text{Before}\\ \text{optimization}}$},{$\substack{\text{After}\\ \text{optimization}}$}},
ymin=0,
ymax=70,
ytick={0,10,20,30,40,50,60},
ylabel style={font=\color{white!15!black}, font=\huge},
ylabel={$R_{\text{sum}}\text{ [bpcu]}$},
axis background/.style={fill=white},
xmajorgrids,
ymajorgrids,
legend style={at={(0.1,0.65)}, anchor=south west, legend cell align=left, align=left, draw=white!15!black, font=\Huge, row sep=0.5em,minimum height=3em}
]

\addplot[black,fill=mycolor1, draw=black] coordinates {
(1,	15.109344514549555) (2,66.277674335954050) 
  };
\addlegendentry{$\substack{\text{SR based on}\\ \text{est. channel}}$}

\addplot[black,fill=mycolor2, draw=black] coordinates {
(1,	15.148854502137443) (2,2.771736283719329) 
  };
\addlegendentry{$\substack{\text{SR based on}\\ \text{true channel}}$}

\end{axis}

\end{tikzpicture}%
}
\caption{Optimization w/ IWMMSE}
\label{fig:Obj_vs_True_IWMMSE}
\end{minipage}
\begin{minipage}{0.48\columnwidth}
 \centering
 \scalebox{0.4}{ 
%
%
\definecolor{mycolor1}{rgb}{0.00000,0.44700,0.74100}%
\definecolor{mycolor2}{rgb}{0.85000,0.32500,0.09800}%
\begin{tikzpicture}

\begin{axis}[%
width=3.521in,
height=3in,
at={(0.758in,0.481in)},
scale only axis,
ybar=2.5pt,
bar width=.8cm,
xmin=0.514285714285714,
xmax=2.48571428571429,
xtick={1,2},
yticklabel style={font=\Large},
xticklabel style={font=\Huge},
xticklabels={{$\substack{\text{Before}\\ \text{optimization}}$},{$\substack{\text{After}\\ \text{optimization}}$}},
ymin=0,
ymax=51,
ytick={0,10,20,30,40,50},
ylabel style={font=\color{white!15!black}, font=\huge},
ylabel={$R_{\text{sum}}\text{ [bpcu]}$},
axis background/.style={fill=white},
xmajorgrids,
ymajorgrids,
legend style={at={(0.1,0.65)}, anchor=south west, legend cell align=left, align=left, draw=white!15!black, font=\Huge, row sep=0.5em,minimum height=3em}
]

\addplot[black,fill=mycolor1, draw=black] coordinates {
(1,	15.109344514549555) (2,49.975096874911200) 
  };
\addlegendentry{$\substack{\text{SR based on}\\ \text{est. channel}}$}

\addplot[black,fill=mycolor2, draw=black] coordinates {
(1,	15.148854502137443) (2,46.349101327235104) 
  };
\addlegendentry{$\substack{\text{SR based on}\\ \text{true channel}}$}

\end{axis}

\end{tikzpicture}%
}
\caption{Optimization w/ AWAMSE}
 \label{fig:Obj_vs_True_LB}
\end{minipage}
\end{figure}
On the contrary, the precoder optimization using the proposed AWAMSE algorithm yields not only an improvement of the SR evaluated using the estimated channel but also of the actual achievable SR, as can be seen in Fig.~\ref{fig:Obj_vs_True_LB}. The results are obtained for the same previous setup with the same instance of the channel.

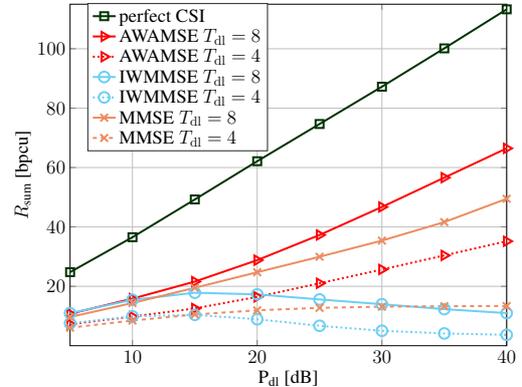
\begin{figure}[h!]
	\centering
	\scalebox{0.5}{
%
%
\definecolor{mycolor1}{rgb}{0.93,0.55,0.38}%
\definecolor{mycolor2}{rgb}{0.59,0.05,0.05}
\begin{tikzpicture}

\begin{axis}[%
width=4.521in,
height=3.566in,
at={(0.758in,0.481in)},
scale only axis,
xmin=5,
xmax=40,
xlabel style={font=\color{white!15!black},font=\Large},
xlabel={$\text{P}_{\text{dl}}\text{ [dB]}$},
ymin=0,
ymax=115,
ylabel style={font=\color{white!15!black}, font=\Large},
ylabel={$R_\text{sum}$ [bpcu]},
axis background/.style={fill=white},
xtick={10,20,30,40},
ytick={20,40,60,80,100},
xmajorgrids,
ymajorgrids,
yticklabel style={font=\Large},
xticklabel style={font=\Large},
legend style={at={(0.04,0.57)}, anchor=south west, legend cell align=left, align=left, draw=white!15!black, font=\Large}
]
%
%
%
%

\addplot [color=green!20!black, line width=1.5pt, mark=square, mark options={solid},mark size=3pt]
  table[row sep=crcr]{%
5	24.775909138478250\\
10	36.535715983448554\\
15	49.253163608006880\\
20	62.122265370657196\\
25	74.659944759345960\\
30	87.219815298528420\\
35	1.001291083788665e+02\\
40	1.132752090156733e+02\\
};
\addlegendentry{perfect CSI}

\addplot [color=red, line width=1.5pt, mark=triangle, mark options={solid, rotate=270, red},mark size=4pt]
  table[row sep=crcr]{%
5	10.755804421906218\\
10  15.851683241100018\\
15  21.591750912937798\\
20	28.806821633313675\\
25  37.293270172719140\\
30	46.723037759383510\\
35	56.596922934008860\\
40  66.460369152523380\\
};
\addlegendentry{AWAMSE $T_\tdl=8$}


\addplot [color=red, dotted, line width=1.5pt, mark=triangle, mark options={solid, rotate=270, red},mark size=4pt]
  table[row sep=crcr]{%
5	7.251933870038728\\
10	9.805825470037606\\
15	12.551973908626387\\
20	16.447111845367328\\
25  21.028657232018210\\
30	25.674778038589400\\
35  30.404900259998770\\
40	35.174417860498580\\
};
\addlegendentry{AWAMSE $T_\tdl=4$}

\addplot [color=cyan!50!white,   line width=1.5pt, mark=o, mark options={solid},mark size=4pt]
  table[row sep=crcr]{%
5	10.947337492378761\\
10	15.464121292954191\\
15	17.839338994122652\\
20	17.290378725120870\\
25	15.584321421640256\\
30	13.973779523861658\\
35	12.314861807016761\\
40	10.983798815447430\\
};
\addlegendentry{IWMMSE $T_\tdl=8$}

%

\addplot [color=cyan!50!white, dotted, line width=1.5pt, mark=o, mark options={solid},mark size=4pt]
  table[row sep=crcr]{%
5	7.538564802994889\\
10	9.886482347491958\\
15  10.393811429474810\\
20  8.911733956587160\\
25	6.704001156771946\\
30	5.032709108714184\\
35	4.134803775782368\\
40  3.657322899602136\\
};
\addlegendentry{IWMMSE $T_\tdl=4$}

\addplot [color=mycolor1,  line width=1.5pt, mark=x, mark options={solid, mycolor1},mark size=4pt]
  table[row sep=crcr]{%
5	9.661232845237075\\
10	14.383320948583437\\
15  19.505703729474206\\
20	24.745937485279110\\
25  29.961011630605004\\
30  35.3931574036468460\\
35	41.630648421536250\\
40  49.482427904699690\\
};
\addlegendentry{MMSE $T_\tdl=8$}


\addplot [color=mycolor1, dashed, line width=1.5pt, mark=x, mark options={solid, mycolor1},mark size=4pt]
  table[row sep=crcr]{%
5	6.067246280352533\\
10	8.470099921783890\\
15	10.509984607801650\\
20	11.925523694343243\\
25	12.747930749689521\\
30	13.141900194092862\\
35	13.305788864776964\\
40	13.363266470801296\\
};
\addlegendentry{MMSE $T_\tdl=4$}

\end{axis}

\end{tikzpicture}%
}
	\caption{Achievable SR versus $P_\tdl$ for $M=32$ antennas, $K=8$ users and different numbers of pilots using AWAMSE and standard IWMMSE}
	\label{fig:SR_LB_IWMMSE}
\end{figure}

In Fig.~\ref{fig:SR_LB_IWMMSE}, we plot the achievable SR versus the transmit power $P_\tdl$ for $M=32$ antennas, $K=8$ users and different numbers of pilots.
One can see that AWAMSE and the standard IWMMSE exhibit comparable performance at low powers, and the gap increases starting from $10$~dB. Due to ignoring the CSI errors during the precoder optimization, the standard IWMMSE fails to mitigate the inter-user interference, especially at high transmit powers, and the SR even degrades in this regime. On the contrary, the AWAMSE shows a remarkably good performance in the medium to high power region and exhibits a high-SNR slope of $1.98$ for $T_\tdl=8$ and $1.12$ for $T_\tdl=4$ compared to $2.63$ for perfect CSI. The latter is obtained by running the standard IWMMSE using the true channel. \\
Interestingly, the MMSE precoder achieves a high SNR slope of $1.71$ in the case $T_\tdl=8=K$. However, for the case $T_\tdl=4<K$, it exhibits a saturation of the achievable SR and fails, therefore, to mitigate the inter-user interference.

\begin{figure}[h!]
	\centering
	\scalebox{0.5}{
%
%
\definecolor{mycolor1}{rgb}{0.00000,0.44700,0.74100}%
\definecolor{mycolor2}{rgb}{0.85000,0.32500,0.09800}%
\begin{tikzpicture}

\begin{axis}[%
width=4.521in,
height=3.566in,
at={(0.758in,0.481in)},
scale only axis,
ybar=2.5pt,
bar width=.4cm,
xmin=0.514285714285714,
xmax=8.48571428571429,
xtick={1,2,3,4,5,6,7,8},
yticklabel style={font=\Large},
xticklabel style={font=\Large},
xticklabels={1,2,3,4,5,6,7,8},
ymin=0,
ymax=.5,
ytick={0,.15,.35,.5},
yticklabels={0, 15, 35,50},
ylabel style={font=\color{white!15!black}, font=\huge},
ylabel={User's power/$P_\tdl$ [$\%$]},
xlabel style={font=\color{white!15!black}, font=\huge},
xlabel={User's index},
axis background/.style={fill=white},
xmajorgrids,
ymajorgrids,
legend style={at={(0.1,0.826)}, anchor=south west, legend cell align=left, align=left, draw=white!15!black, font=\LARGE}
]

\addplot[black,fill=mycolor1, draw=black] coordinates {
(1,0.125028679223838) (2,0) (3, 0.318398740560628) (4,0.360079860058411) (5,0) (6,0) (7,0) (8,0.196492720157123)
  };
\addlegendentry{AWAMSE}

\addplot[black,fill=mycolor2, draw=black] coordinates {
(1,	0.117052339371312) (2,5.947669512389248e-86) (3,0.300267367542695) (4,0.387214096161314)(5,0) (6,0) (7,0) (8,0.195466196924680)
  };
\addlegendentry{SIWMMSE}

\end{axis}

\end{tikzpicture}%
}
	\caption{Power allocation among the users for $P_\tdl=40$~dB}
	\label{fig:PowerPercentage}
\end{figure}
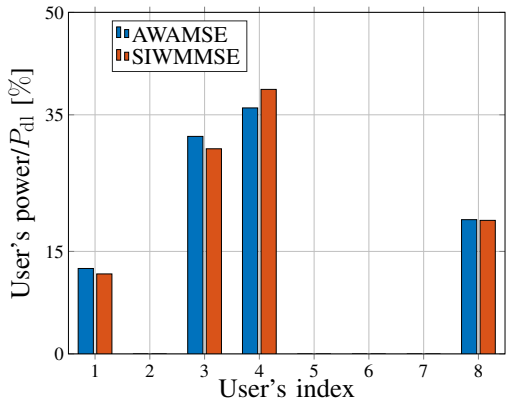

Next, we compare the performance of the proposed AWAMSE algorithm to the adapted SIWMMSE approach presented in Section~\ref{sec:sIWMMSE} for two setups with different pilot numbers, namely $T_\tdl=K$ and $T_\tdl<K$. For the SIWMMSE, we generate $N=100$~samples according to \eqref{eq:sampleGen}.
\begin{figure}[h!]
	\centering
	\scalebox{0.5}{
%
%
\definecolor{mycolor1}{rgb}{0.07451,0.62353,1.00000}%
\begin{tikzpicture}

\begin{axis}[%
width=4.521in,
height=3.566in,
at={(0.758in,0.481in)},
scale only axis,
xmin=5,
xmax=40,
xlabel style={font=\color{white!15!black},font=\Large},
xlabel={$\text{P}_{\text{dl}}\text{ [dB]}$},
ymin=0,
ymax=75,
ylabel style={font=\color{white!15!black}, font=\Large},
ylabel={$R_\text{sum}$ [bpcu]},
axis background/.style={fill=white},
xtick={10,20,30,40},
ytick={0,20,40,60},
xmajorgrids,
ymajorgrids,
yticklabel style={font=\Large},
xticklabel style={font=\Large},
legend style={at={(0.15,0.7)}, anchor=south west, legend cell align=left, align=left, draw=white!15!black, font=\Large}
]

\addplot [color=mycolor1, line width=1.5pt, mark=o, mark options={solid, mycolor1},mark size=4pt]
  table[row sep=crcr]{%
5	11.305044779808464\\
10  17.307024145042316\\
15	24.359118230692857\\
20	32.349163916716705\\
25	41.048947112996100\\
30	50.171005157337030\\
35  59.354421783764990\\
40	68.326375810320800\\
};
\addlegendentry{SIWMMSE}

\addplot [color=red, line width=1.5pt, mark=triangle, mark options={solid, rotate=270, red},mark size=4pt]
  table[row sep=crcr]{%
5	10.755804421906218\\
10  15.851683241100018\\
15  21.591750912937798\\
20	28.806821633313675\\
25  37.293270172719140\\
30	46.723037759383510\\
35	56.596922934008860\\
40  66.460369152523380\\
};
\addlegendentry{AWAMSE}



\addplot [color=mycolor1, dashed, line width=1.5pt, mark=o, mark options={solid, mycolor1},mark size=4pt]
  table[row sep=crcr]{%
5	8.352537809557031\\
10	12.480960473079914\\
15	17.103472114905607\\
20	22.285304640568327\\
25  27.916977735724934\\
30	33.779397589274126\\
35	39.841522787529620\\
40	45.9368999297529540\\
};

\addplot [color=red, dashed, line width=1.5pt, mark=triangle, mark options={solid, rotate=270, red},mark size=4pt]
  table[row sep=crcr]{%
5	7.251933870038728\\
10	9.805825470037606\\
15	12.551973908626387\\
20	16.447111845367328\\
25  21.028657232018210\\
30	25.674778038589400\\
35  30.404900259998770\\
40	35.174417860498580\\
};

\end{axis}

\draw [black, very thick] (11.75,8.25) ellipse [x radius=0.3, y radius=0.5];
\draw [black, very thick] (11.75,5.5) ellipse [x radius=0.3, y radius=.9];

\draw [-stealth, thick](11.25,9.6) -- (11.65,8.7);
\node at (11,9.7) {\Large $T_\text{dl}=8$};
\draw [-stealth, thick](11.25,3.6) -- (11.65,4.6);
\node at (11,3.5) {\Large $T_\text{dl}=4$};

\end{tikzpicture}%
}
	\caption{Achievable SR versus $P_\tdl$ for $M=32$ antennas, $K=8$ users and different numbers of pilots using AWAMSE and SIWMMSE}
	\label{fig:SR_LB_SIWMMSE}
\end{figure}
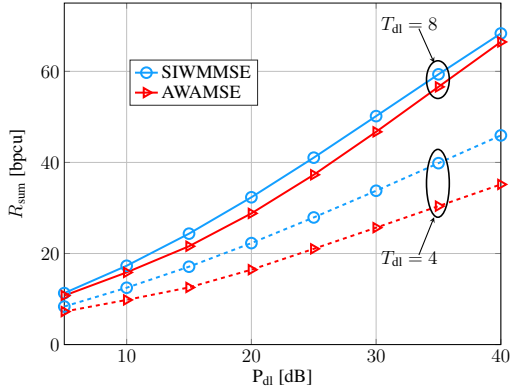
One can observe in Fig.~\ref{fig:SR_LB_SIWMMSE} that both algorithms exhibit approximately the same high SNR slope for both cases. While the proposed AWAMSE algorithm achieves comparable performance in terms of the SR as SIWMMSE for $T_\tdl=8$, the gap between both methods increases for $T_\tdl=4$. 
Note that we specifically investigate the cases where $T_\tdl=K$ and $T_\tdl<K$, since, as shown in our recent work \cite{ZF_LS}, an interference-free transmission among the users can be guaranteed if $T_\tdl\geq K$ for asymptotically high transmit power. This means if this condition is not satisfied, optimizing the transmit strategy would lead to a maximum of $T_\tdl$ active users. This can be confirmed by considering the case $T_\tdl=4$ and evaluating the ratio of the users' powers to the total transmit power after running both algorithms as shown in Fig.\ref{fig:PowerPercentage}. It can seen that for both algorithms, users 2, 5, 6, and 7 are inactive, and therefore, only $T_\tdl=4$ users are served. \\
Although SIWMMSE outperforms AWAMSE in terms of the achievable SR, the latter has the advantage that no Monte Carlo runs are needed inside each iteration to average over all variables for all channel realizations before updating the precoding matrix and is, therefore, much faster. At $40$~dB, for example, the AWAMSE algorithm exhibits a run time that is approximately ten times smaller than that of the SIWMMSE.

	\section{Conclusion}
	We have derived a lower bound on the achievable downlink SINR that takes into account the incomplete channel knowledge available at the transmitter. Connecting this lower bound to an average MSE allowed us to design an efficient precoding strategy where all updates are given in closed form. Even in challenging scenarios with too few DL pilots, our approach is competitive with the more computationally demanding SIWMMSE.
	
	\section{Appendix}\label{Appendix}
The first derivative of \eqref{eq:AWAMSEmin3} w.r.t. $\delta$ reads as
\begin{equation}
	\begin{aligned}
		\frac{\partial}{\partial \delta}&=2\tr(\bs{A}^{-1}\bs{T}^\HH \bs{U}\bs{U}\bs{T} \bs{A}^{-1})-2\tr(\bs{B}^{-1}\bs{A}^{-1}\bs{T}^\HH \bs{U}\bs{U}\bs{T} \bs{A}^{-1})\\
		&-2\frac{1}{P_\tdl} \tr(\bs{A}^{-2}\bs{T}^\HH \bs{U}\bs{U}\bs{T} \bs{A}^{-1})\tr (\bs{U}\bs{G}\bs{G}^\ast)
	\end{aligned}
\end{equation}
where $\bs{B}^{-1}=\bs{A}^{-1}\bs{X}=(\bs{I}+\delta\bs{X}^{-1})^{-1}$ and $\bs{X}=\bs{Z}+\bs{T}^\HH\bs{U}\bs{T}$. From the matrix inversion lemma (e.g. \cite{Scharf1989}), it follows that $\bs{I}-\bs{B}^{-1}=\delta \bs{A}^{-1}$. The derivative can therefore be rewritten as
\begin{equation}
	\begin{aligned}
		\frac{\partial}{\partial \delta}&=2\tr(\bs{A}^{-2}\bs{T}^\HH \bs{U}\bs{U}\bs{T} \bs{A}^{-1}) \left(\delta -\frac{\tr (\bs{U}\bs{G}\bs{G}^\ast)}{P_\tdl}\right)\overset{!}{=} 0
	\end{aligned}
\end{equation}
and the optimal $\delta^\text{opt}$ is hence given by \eqref{eq:DeltaOpt}.

\end{document}